\documentclass[12pt]{article}

\usepackage[dvipdfmx]{graphicx}
\usepackage{amsfonts}
\usepackage[mathscr]{eucal}
\usepackage{ascmac}
\usepackage{dcolumn}
\usepackage{bm}
\usepackage[colorlinks=true,linkcolor=blue,citecolor=blue]{hyperref}
\usepackage{color}

\begin{document}


\newcommand{\gtrsim}{ \mathop{}_{\textstyle \sim}^{\textstyle >} }
\newcommand{\lesssim}{ \mathop{}_{\textstyle \sim}^{\textstyle <} }
\newcommand{\vev}[1]{ \left\langle {#1} \right\rangle }
\newcommand{\bra}[1]{ \langle {#1} | }
\newcommand{\ket}[1]{ | {#1} \rangle }
\newcommand{\EV}{ \ {\rm eV} }
\newcommand{\KEV}{ \ {\rm keV} }
\newcommand{\MEV}{\  {\rm MeV} }
\newcommand{\GEV}{\  {\rm GeV} }
\newcommand{\TEV}{\  {\rm TeV} }
\newcommand{\1}{\mbox{1}\hspace{-0.25em}\mbox{l}}
\newcommand{\Red}[1]{{\color{red} {#1}}}

\newcommand{\lmk}{\left(}  
\newcommand{\rmk}{\right)}
\newcommand{\lkk}{\left[}  
\newcommand{\rkk}{\right]}
\newcommand{\lhk}{\left \{ }  
\newcommand{\rhk}{\right \} }
\newcommand{\del}{\partial}  
\newcommand{\la}{\left\langle} 
\newcommand{\ra}{\right\rangle}
\newcommand{\half}{\frac{1}{2}}

\newcommand{\bea}{\begin{array}}
\newcommand{\eea}{\end{array}}
\newcommand{\beq}{\begin{eqnarray}}
\newcommand{\eeq}{\end{eqnarray}}

\newcommand{\dd}{\mathrm{d}}
\newcommand{\Mpl}{M_{\rm Pl}}
\newcommand{\mg}{m_{3/2}}
\newcommand{\abs}[1]{\left\vert {#1} \right\vert}
\newcommand{\mphi}{m_{\phi}}
\newcommand{\Hz}{\ {\rm Hz}}
\newcommand{\for}{\quad \text{for }}
\newcommand{\Min}{\text{Min}}
\newcommand{\Max}{\text{Max}}
\newcommand{\Kahler}{K\"{a}hler }
\newcommand{\cphi}{\varphi}

\begin{titlepage}

\baselineskip 8mm

\begin{flushright}
ICRR-Report-698-2014-24 \\
IPMU 15-0017
\end{flushright}

\begin{center}

\vskip 1.2cm

{\Large\bf
Affleck-Dine baryogenesis after D-term inflation 
and~solutions~to~the~baryon-DM~coincidence~problem 
}

\vskip 1.8cm

{\large Masahiro Kawasaki$^{a,b}$
and
Masaki Yamada$^{a,b}$}

\vskip 0.4cm

{\it$^a$Institute for Cosmic Ray Research, The University of Tokyo,
5-1-5 Kashiwanoha, Kashiwa, Chiba 277-8582, Japan}\\
{\it$^b$Kavli IPMU (WPI), TODIAS, The University of Tokyo, 5-1-5 Kashiwanoha, 
Kashiwa, 277-8583, Japan}

\date{\today}
\vspace{2cm}

\begin{abstract}  
We investigate the Affleck-Dine baryogenesis  after D-term
inflation  with a positive Hubble-induced mass term for a $B-L$ flat
direction. It stays at a large field value during D-term inflation,  
and  just after
inflation ends it starts to oscillate around the origin of the potential  due
to the positive Hubble-induced mass term.  The phase direction is kicked by
higher-dimensional \Kahler potentials  to generate the $B-L$ asymmetry.  The
scenario predicts nonzero baryonic isocurvature perturbations,  which would be
detected by future observations of CMB fluctuations.  We also provide a D-term
inflation model which naturally explain the coincidence of the energy density
of baryon and dark matter. 
\end{abstract}


\end{center}
\end{titlepage}

\baselineskip 6mm


\section{Introduction
\label{sec:introduction}}

Inflation is a new paradigm to solve cosmological problems 
related to the initial conditions of the early Universe. 
However, any preexisting baryon asymmetry is diluted away 
due to the inflationary expansion of the Universe, 
so that there should be a mechanism to generate the observed baryon 
asymmetry after inflation. 
In addition, the observed abundance of baryon asymmetry 
is equal to that of dark matter (DM) within of order unity 
($\Omega_b / \Omega_{\rm DM} \simeq 0.2$)~\cite{pdg}, 
which is a mystery in cosmology referred to as the baryon-DM coincidence 
problem. 
This implies that the baryon asymmetry and DM have a common origin. 
We therefore need to consider inflation, baryogenesis, and DM production 
simultaneously to construct consistent cosmological models.

In this paper, we focus on 
D-term inflation models by the following reasons~\cite{Binetruy:1996xj, Halyo:1996pp}. 
First, the energy scale of D-term inflation is naturally of order the GUT 
scale, which predicts an amplitude of CMB fluctuations consistent with the 
observations~\cite{Hinshaw:2012aka, Ade:2013zuv}. 
The second reason is related to the so-called $\eta$-problem. 
If inflation is driven by a nonzero F-term potential energy, 
supergravity effects induce masses of order the Hubble parameter
to all scalar fields, including inflaton. 
However, 
the Hubble-induced mass for inflaton 
spoils the flatness of its potential 
and results in an $\mathcal{O}(1)$ slow roll parameter $\eta \sim 1$. 
Since Hubble-induced masses come only from nonzero F-term potential energy, 
the $\eta$ problem is absence in the case of D-term inflation, 
in which inflation is driven by nonzero D-term potential energy.

In supersymmetric (SUSY) theories, 
baryon asymmetry can be generated by the Affleck-Dine baryogenesis 
using a $B-L$ charged flat direction called an AD field~\cite{AD, DRT}. 
Since the soft mass of the AD field is much smaller than the energy scale of 
inflation, it can stay at a large vacuum expectation value (VEV) 
during D-term inflation. 
After D-term inflation ends, the AD field obtains a Hubble-induced mass 
from the oscillating inflaton field through the supergravity effects. 
In the literature, 
the Hubble-induced mass term was assumed to be negative 
(i.e., tachyonic) or absent to investigate the dynamics of the AD field and calculate 
the produced baryon asymmetry~\cite{Kolda:1998kc, Enqvist:1998pf, Enqvist:1999hv, Kawasaki:2001in}. 
In this paper, we investigate the Affleck-Dine baryogenesis 
in the case that the AD field obtains a positive Hubble-induced mass term 
after D-term inflation. 
In this case, the AD field starts to oscillate around the origin of the 
potential due to the positive Hubble-induced mass term just after inflation. 
At the same time, the phase direction of the AD field feels 
non-renormalizable $B-L$ breaking operators and starts to rotate in the phase 
space, which results in a generation of $B-L$ asymmetry. 
Then, the coherent oscillation of the AD field decays and dissipates into the 
thermal plasma and the $B-L$ asymmetry is converted to the desired baryon 
asymmetry through the sphaleron effects~\cite{Kuzmin:1985mm, Fukugita:1986hr}. 
We calculate the baryon asymmetry and show that the result can be consistent 
with the observed amount of baryon asymmetry. 
Since the AD field fluctuates 
due to the absence of Hubble-induced mass during D-term inflation, 
the Affleck-Dine baryogenesis predicts some amount of baryonic isocurvature 
perturbations~\cite{Enqvist:1998pf, Enqvist:1999hv, Kawasaki:2001in, Kasuya:2008xp, Harigaya:2014tla}. 
Especially, when the AD field starts to oscillate due to the positive 
Hubble-induced mass term, its radial direction also has quantum fluctuations 
and contributes to the baryonic isocurvature perurbations. 
If one might consider a high-scale D-term inflation model, the resulting 
isocurvature perturbations would be detected by future observations of CMB 
fluctuations.

We also build a D-term inflation model which naturally predicts 
the baryon-to-DM ratio of order unity. 
We introduce a shift symmetry for the inflaton superfield to ensure a large 
initial VEV of inflaton~\cite{Kawasaki:2000yn}. 
We also introduce its linear term in the \Kahler potential 
so that the branching of inflaton decay into gravitinos can be of 
order unity~\cite{
Nakamura:2006uc, Kawasaki:2006gs, Asaka:2006bv, Dine:2006ii, Endo:2006tf, Kawasaki:2006hm, Endo:2006qk}. 
When the mass of gravitino is larger than O(100) TeV, 
it decays into the minimal SUSY standard model (MSSM) particles before the 
big bang nucleosynthesis (BBN) epoch. 
The DM, which is the lightest SUSY particle (LSP), is therefore produced 
non-thermally from the gravitino decay. 
This scenario, together with the above Affleck-Dine baryogenesis scenario, 
predicts an $\mathcal{O}(1)$ ratio of the energy density of baryon and DM. 
This means that the scenario naturally explains the baryon-DM coincidence 
problem. 
This arises from the fact that both of them is related to the energy scale 
of inflation. 
The amount of baryon asymmetry is proportional to the reheating temperature 
of the Universe and inversely proportional to the Hubble parameter during 
inflation. 
The DM abundance is proportional to the reheating temperature 
and inversely proportional to the mass of inflaton. 
Since the Hubble parameter and the mass is related to each other, 
the resulting baryon and DM density is naturally of order unity. 
We predict that the LSP mass is 
two orders of magnitude larger than the proton mass, 
which comes from the fact that 
the GUT scale is two orders of magnitude less than the Planck scale. 
When the LSP is mostly wino or higgsino, 
it would be detected by future indirect detection experiments of DM.

This paper is organized as follows.
In the next section, we briefly review the simplest D-term inflation model 
as an illustration. 
In Sec.~\ref{sec:ADBG}, 
we consider the Affleck-Dine baryogenesis and calculate the baryon asymmetry 
and baryonic isocurvature fluctuations 
in the case that the AD field obtains 
a positive Hubble-induced mass term after inflation. 
In Sec.~\ref{sec:coincidence}, 
we provide a D-term inflation model 
which naturally predicts the baryon-to-DM ratio of order unity. 
Section~\ref{sub:summary} is devoted to the summary.

\section{D-term inflation
\label{sec:D-term inflation}}

We focus on D-term inflation~\cite{Binetruy:1996xj, Halyo:1996pp}, in which 
inflation is driven by a finite energy density of the D-term potential. 
Although the following simple model of D-term inflation predicts 
the spectral index relatively blue tilted 
compared with the observation of CMB fluctuations, 
we review it as an illustration. 
Note that there are variants of D-term inflation models 
which predict the spectral index consistent with the observed value~\cite{Buchmuller:2012ex, Buchmuller:2013zfa}, 
and the results in the next section can be applied to those models, too.%
\footnote{
The results are not applicable to the inflation model considered 
in Ref.~\cite{Buchmuller:2014rfa} because a F-term potential drives inflation 
with a sizable e-folding number in that scenario. 
}

We introduce a $U(1)$ gauge symmetry with a Fayet-Ilipoulos (FI) term $\xi$ 
and consider superfields $S$, $\psi_-$, and $\psi_+$ 
with $U(1)$ gauge charges as $0$, $-1$, and $1$, respectively.%
\footnote{%
The FI term can be generated dynamically as discussed 
in Ref.~\cite{Domcke:2014zqa}. 
They also argue that the production of cosmic strings at the end of inflation 
is generally avoided when the FI term is generated dynamically. 
Otherwise 
the CMB data puts an upper bound on a cosmic string contribution to the CMB 
fluctuations, which leads to the upper bound on 
the FI term~\cite{Buchmuller:2012ex, Battye:2010hg}. 
}
The D-term potential is written as 
\beq
 V_D = \frac{g^2}{2} \lmk \abs{\psi_+}^2 - \abs{\psi_-}^2 - \xi \rmk^2, 
\eeq
where $g$ is the $U(1)$ gauge coupling constant. 
We introduce a superpotential given as 
\beq
 W^{\rm (inf)} = \lambda S \psi_+ \psi_-, 
 \label{W_inf}
\eeq
where $\lambda$ is a coupling constant. 
Hereafter, we denote their scalar components by the same symbols as 
the superfields.

The scalar component of the field $S$ plays a role of inflaton. 
Suppose that the inflaton $S$ has a VEV larger than the critical value of 
$S_c \equiv g \sqrt{\xi} / \lambda$. 
The fields $\psi_-$ and $\psi_+$ obtain large effective masses from the VEV 
of the inflaton and stays at the origin of the potential. 
In this regime, the nonzero D-term potential of $V_0 = g^2 \xi^2/2$ drives 
inflation. 
The Coleman-Weinberg potential for the inflaton lifts its potential 
above the critical point such as 
\beq
 V_{\rm 1-loop} 
 \simeq
 \half g^2 \xi^2 \lmk 1+ 
 \frac{g^2}{16 \pi^2} 
 \log \frac{\lambda^2 \abs{S}^2}{Q^2} \rmk, 
 \label{CW potential}
\eeq
where $Q$ is a renormalization scale. 
Thus, the inflaton slowly rolls down to the origin of the potential. 
The COBE normalization requires~\cite{Hinshaw:2012aka, Ade:2013zuv}
\beq
 \sqrt{\xi} \simeq 6.6 \times 10^{15} \GEV. 
\eeq
This leads to the Hubble parameter during inflation such as 
\beq
 H_I \simeq \frac{g \xi/\sqrt{2}}{\sqrt{3} \Mpl} 
 \simeq 3.7 \times 10^{12} \GEV \lmk \frac{\sqrt{\xi}}{6.6 \times 10^{15} \GEV} \rmk^2. 
\eeq
The e-folding number and a slow roll parameter are calculated as 
\beq
 N_* &\simeq& \frac{4 \pi^2}{g^2} \frac{S_*^2}{\Mpl^2}, \\
 \eta &\equiv& \frac{V''}{V} \Mpl^2 
 \simeq - \frac{g^2}{8 \pi^2} \frac{\Mpl^2}{S_*^2} \simeq - \frac{1}{2 N_*},
\eeq
where the subscript $_{*}$ denotes values corresponding to the pivot scale
$k_* = 0.05$Mpc$^{-1}$.
The slow roll condition fails ($\eta \sim 1$) at the VEV around 
$S \simeq g/(2 \sqrt{2} \pi)$, which is larger than the critical value 
$S_c$ for the case of $\lambda = \mathcal{O}(1)$. 
Thus, 
slow roll inflation ends at the VEV around $S \simeq g/(2 \sqrt{2} \pi)$ 
and soon after that 
the waterfall field $\psi_+$ starts to oscillate around the low energy 
minimum of $\sqrt{\xi}$.

The scalar spectral index is calculated as 
\beq
 n_s \simeq \frac{1}{N_*} \simeq 0.98. 
\eeq
It is measured by $Planck$ data alone such as~\cite{Ade:2013zuv}
\beq
 n_s^{(obs)} = 0.9616 \pm 0.0094 \quad \lmk 68\% \rmk. 
\eeq
So the above prediction deviates by about $2 \sigma$ from the observation. 
Let us emphasize that 
the results in the next section can be applied to 
other variants of D-term inflation models, 
including the ones 
which predict the spectral index consistent with the observed value 
within a $1 \sigma$ level~\cite{Buchmuller:2012ex, Buchmuller:2013zfa}.

After inflation ends, 
the energy density of the Universe is dominated by that of the oscillation of 
$S$ and $\psi_+$. 
When some MSSM fields carry nonzero $U(1)$ charge, 
the field $\psi_+$ immediately decays into the MSSM fields through 
the interaction in the D-term potential. 
Even if the MSSM fields have no $U(1)$ charge, 
the kinetic mixings between the $U(1)$ and $U(1)_Y$ 
makes the field $\psi_+$ decay into the MSSM fields relatively fast~\cite{Kolda:1998kc}. 
Thus, 
the reheating temperature of the Universe is determined by the relatively 
late-time decay of the inflaton $S$, which dilutes the relics produced from 
the decay of $\psi_+$. 
We define the reheating temperature 
as 
\beq
 T_{\rm RH} 
 &\simeq& \lmk \frac{90}{g_* (T_{\rm RH}) \pi^2} \rmk^{1/4} 
 \sqrt{\Gamma_S \Mpl}, 
 \label{T_RH}
\eeq
where 
$\Gamma_S$ is the decay rate of the inflaton $S$. 
The reheating temperature of the Universe depends on the mass of the inflaton 
($m_S \equiv \lambda \sqrt{\xi}$) 
and the assumption of interactions between $S$ and the MSSM fields, 
which is determined by their underlying symmetry. 
We explicitly calculate the reheating temperature 
in Sec.~\ref{sec:coincidence} 
for a specific model with a shift symmetry and an approximate $Z_2$ symmetry, 
while we regard it as a free parameter in the next section 
to calculate the the amount of baryon asymmetry generated by the 
Affleck-Dine baryogenesis.

\section{Affleck-Dine baryogenesis
\label{sec:ADBG}}

In this section, we consider the Affleck-Dine baryogenesis~\cite{AD,DRT}  
after D-term inflation  and calculate the resulting baryon
asymmetry and baryonic isocurvature perturbations.  
We consider the case that  the AD field obtains a positive Hubble-induced 
mass term after the end of inflation,  which has been overlooked in the
literature.  
We investigate the potential of the AD field  after
D-term inflation in the next subsection,  and then we calculate the
baryon asymmetry.  
In Sec.~\ref{subsec:isocurvature},  we calculate baryonic isocurvature 
perturbations predicted by the Affleck-Dine baryogenesis.  
Finally, we comment on Q-ball formation in Sec.~\ref{subsec:Q-ball}.

\subsection{Potential of the AD field
\label{subsec:potential}}

We consider the Affleck-Dine baryogenesis~\cite{AD, DRT} 
using a flat direction ($=$AD field) $\phi$ with nonzero $B-L$ charge. 
The AD field has soft SUSY breaking terms 
through the low-energy SUSY breaking effect. 
Since the soft mass of the flat direction is much smaller than the Hubble parameter 
during inflation, 
it has a large VEV during inflation. 
In this paper, we assume that the superpotential of the AD field is absent or 
sufficiently small so that 
the initial VEV of the AD field $\phi_i$ 
can be as large as the Planck scale ($\phi_i \simeq \Mpl$).%
\footnote{
For example, we can introduce R-symmetry 
in which the charge of the AD field is zero 
to forbid the superpotential for the AD field.
This symmetry is not exact 
because it is inconsistent with 
the constant term in the superpotential which is needed to ensure the (almost) vanishing cosmological constant 
as well as the gaugino mass terms. 
Since the R-symmetry breaking order parameter is of order the gravitino mass, 
which is much smaller than the energy scale of inflation, 
such breaking terms can be neglected in the following discussion. 
}
Such a large VEV is favoured to avoid the baryonic isocurvature constraint as shown in 
Sec.~\ref{subsec:isocurvature} (see also Ref.~\cite{Harigaya:2014tla}). 
Note that owing to the exponential term in the supergravity potential 
the VEV of the AD field is restricted below the Planck scale. 
Since the curvature of the phase direction is absent (or at least much less 
than the Hubble parameter), 
the phase of the flat direction also stays at a certain phase during 
inflation. 
We denote the initial phase of the AD field as $\theta_i$.

After inflation, 
the energy density of the Universe is dominated by that of oscillating 
inflaton. 
Since the inflaton oscillation induces F-term potential, 
the flat direction obtains Hubble-induced terms 
through supergravity effects~\cite{DRT}. 
The scalar potential in supergravity is given as 
\beq
 V = e^{K/\Mpl^2} 
 \lkk \lmk D_i W \rmk K^{i \bar{j}} \lmk D_j W \rmk^* 
 - \frac{3}{\Mpl^2} \abs{W}^2 \rkk, 
\eeq
where $K$ is a \Kahler potential and $D_i W \equiv W_i + K_i W / \Mpl^2$. 
The subscripts represent the derivative with respect to field $i$
and $K^{i \bar{j}} \equiv (K_{i \bar{j}})^{-1}$. 
Since the F-term of $\psi_-$ is given by $\lambda S \psi_+$, 
the scalar potential includes the following term:%
\beq
 V \supset \frac{\abs{\phi}^2}{\Mpl^2} \abs{\lambda S \psi_+ }^2. 
\eeq
Using $\psi_+ = \sqrt{\xi}$ and taking average with respect to time, 
we obtain the Hubble-induced mass term of 
\beq
 V \supset \frac{3}{2} H^2(t) \abs{\phi}^2. 
\eeq
Here we have used the virial theorem: 
\beq
 m_S^2 \la \abs{S}^2 \ra 
 \simeq \frac{3}{2} H^2(t) \Mpl^2, 
\eeq
where $\la \ra$ represents the time-average. 
In the literature, 
they assume a negative Hubble-induced mass term 
which comes from 
higher-dimensional \Kahler potentials, say, 
\beq
 K^{(H)} = 
 c_S \abs{S}^2 
 \frac{\abs{\phi}^2}{\Mpl^2}, 
 \label{K_H}
\eeq
where $c_S$ is an $\mathcal{O}(1)$ constant. 
As we stated in the introduction, we consider the positive Hubble-induced 
mass term and do not need to introduce such higher-dimensional terms. 
Hereafter, we consider a general case 
and denote the coefficient of the Hubble-induced mass term 
as $c_H$ ($= \mathcal{O}(1)$).

In addition to the Hubble-induced mass term, 
the flat direction obtains higher-dimensional terms 
from non-renormalizable \Kahler potentials.%
\footnote{
The usual Hubble-induced A-terms are absent during the inflaton oscillation era. 
}
The following \Kahler potential may exist 
and 
induce $U(1)$ breaking higher-dimensional potential 
for the flat direction: 
\beq
 && \frac{2}{3} a_H \int \dd^2 \theta \dd^2 \bar{\theta} 
 \abs{S}^2 
 \frac{\phi^n}{n \Mpl^n} + c.c.  
 \nonumber \\ 
 &\simeq& - \frac{2}{3} a_H  
 \abs{\del_\mu S}^2 
 \frac{\phi^n}{n \Mpl^n} +c.c. 
 \nonumber \\ 
 &\simeq& - a_H H^2(t) \lmk \frac{\phi^n}{n \Mpl^{n-2}} + c.c. \rmk, 
 \label{a_H}
\eeq
where $n$ is an integer depending on flat directions. 
For example, 
$n=3, 6, 9, \dots$ for $u^c d^c d^c$ flat direction. 
In the last line, we take average with respect to time and 
use the relation of $\la (\del_0 S)^2 \ra \simeq 3 H^2(t) \Mpl / 2$, 
which comes from the virial theorem. 
Note that this term has a nonzero phase which is different from 
the phase of the flat direction $\theta_i$ during inflation. 
We can redefine the phase of the flat direction to eliminate 
the phase of $a_H$. 
After the elimination, we redefine the initial phase of the flat direction 
as $\theta_i$ without loss of generality. 
The discrepancy between the initial phase of the flat direction 
and the phase of the above $U(1)$ breaking term 
is essential to generate the baryon asymmetry.

In summary, 
the AD field obtains the following potential after inflation: 
\beq
 V(\phi) = c_H H^2(t) \abs{\phi}^2 
 - a_H H^2 (t) \lmk \frac{\phi^n}{n \Mpl^{n-2}} + c.c. \rmk 
 + \dots, 
 \label{potential}
\eeq
where $c_H$ and $a_H$ are positive $\mathcal{O}(1)$ parameters. 
The dots represents higher-dimensional terms which restrict the AD field below 
the Planck scale. 
Since the flat direction starts to oscillate due to the Hubble-induced mass, 
we can neglect usual soft mass and A-terms for the AD field.

\subsection{Calculation of baryon asymmetry
\label{subsec:ADBG}}

In this subsection, we calculate the baryon asymmetry generated 
from the AD field 
with the potential~(\ref{potential}).
The initial VEV and phase are 
$\phi_i$ and $\theta_i$, respectively. 
For $c_H >0$, the flat direction starts to oscillate around the origin 
of the potential 
just after the end of inflation. 
At the same time, the flat direction is kicked in the phase direction 
due to the second term in Eq.~(\ref{potential}). 
The $B-L$ asymmetry is generated through this dynamics. 
The evolution of equation for the $B-L$ charge density is written as 
\beq
\dot{n}_{B-L} +3 H n_{B-L} 
=  2 q \,{\rm Im}\lkk \phi^* \frac{\del V}{\del \phi^*} \rkk, 
\eeq
where $q$ denotes the $B-L$ charge of the AD field. 
From this equation we obtain 
\beq
  a^3 n_{B-L} (t_{\rm osc}) 
 &\simeq& 
 \int \dd t \,2 q a^3(t) \abs{\phi V_A'} \sin (n \theta) 
 \nonumber \\
 &\equiv& 
 \epsilon q H_I \phi_i^2 \label{n_B-L}\\
 \epsilon 
 &\simeq& 
 (3-4) \times \frac{8}{3 n - 6} 
 a_H \sin \lmk - n \theta_i \rmk \lmk \frac{\phi_i }{ \Mpl} \rmk^{n-2} 
 \label{epsilon}
\eeq
where we have used $\phi \propto a^{-3/4}$. 
We define $\epsilon$ ($\le 1$) which represents the efficiency of 
baryogenesis. 
We have numerically solved the equations of motion for $\phi$ 
and $S$ with the Friedmann equation 
and have obtained the numerical factor of $(3-4)$ for $\epsilon \lesssim 1$. 
Since the baryon density has to be smaller than the number density of the AD 
field, $\epsilon$ is at most unity even for large $a_H$ and $\phi_i$. 
The amplitude of the flat direction decreases as time evolves 
due to the Hubble expansion 
and the $B-L$ breaking effect is absent soon after the oscillation. 
Thus, the generated $B-L$ asymmetry is conserved soon after the AD field 
starts to oscillate.

Then, the oscillating AD field decays and dissipates into 
radiation~\cite{Mukaida:2012qn} and the sphaleron effect converts 
the $B-L$ asymmetry to the baryon asymmetry~\cite{Kuzmin:1985mm, Fukugita:1986hr}. 
Since the sphaleron process is in thermal equilibrium, 
the resultant baryon asymmetry is related to the $B-L$ asymmetry 
as~\cite{Harvey:1990qw}
\beq
 n_b = \frac{8}{23} 
 n_{B-L}. 
\eeq
Assuming the absence of entropy production other than the reheating by 
inflaton decay, 
we can calculate the resulting baryon-to-entropy ratio $Y_b$ as 
\beq
 Y_b  
 &\equiv& 
 \frac{n_b}{s} 
 = 
 \left. \frac{8}{23} \frac{n_{B-L}}{s} \right\vert_{\rm RH} 
 \simeq
 \left. \frac{8}{23} 
 \frac{3 T_{\rm RH} n_{B-L}}{4 \rho_S} \right\vert_{\rm osc} 
 \simeq 
 \frac{8}{23} 
 \frac{\epsilon q T_{\rm RH}}{4 H_I} \lmk \frac{\phi_i}{\Mpl} \rmk^2
 \nonumber\\[0.5em]
 &\simeq& 
 8.7 \times 10^{-11} 
 \epsilon q 
 \lmk \frac{T_{\rm RH}}{4 \times 10^3 \GEV} \rmk 
 \lmk \frac{H_I}{4 \times 10^{12} \GEV} \rmk^{-1} 
 \lmk \frac{\phi_i}{\Mpl} \rmk^2, 
 \label{Y_b}
\eeq
where $\rho_S$ ($\simeq 3 H_I^2 \Mpl^2$) is the energy density of 
the inflaton $S$. 
This can be consistent with the observed baryon asymmetry of 
$Y_b^{\rm obs} \simeq 8.7 \times 10^{-11}$~\cite{pdg}.

You can find 
differences from the conventional scenario of the Affleck-Dine baryogenesis. 
The Hubble parameter at the beginning of oscillation $H_I$ 
is determined by the energy scale of inflation, 
not by the curvature of the potential for the flat direction (see Ref.~\cite{Harigaya:2014tla}, for example). 
This is because the flat direction starts to oscillate just after the end of 
inflation due to the positive Hubble-induced mass term, 
while in the conventional scenario it starts to oscillate at 
$H(t) \simeq m_\phi$, where $m_\phi$ is the soft mass of the AD field. 
This allows us to consider a relatively large reheating temperature 
even if the initial VEV $\phi_i$ is as large as the Planck scale. 
In addition, the ellipticity parameter $\epsilon$, which describes the 
efficiency of baryogenesis, 
can be much smaller than unity when $\phi_i$ is smaller than the Planck scale. 
This is because the phase direction of the AD field is kicked by a 
higher-dimensional \Kahler potential, 
which is highly suppressed for a small VEV of the AD field. 
However, 
as shown in the next subsection, 
the baryonic isocurvature constraint requires the initial VEV to be 
as large as the Planck scale. 
In that case, $\epsilon$ is of order unity for $a_H = \mathcal{O}(1)$.

One might wonder if the energy density of the AD field dominates that of the 
Universe in the case that its initial VEV is as large as the Planck scale. 
This may be true in the case of conventional Affleck-Dine baryogenesis, 
in which the AD field starts to oscillate when the Hubble parameter decreases 
down to the soft mass of the flat direction. 
However, the energy density of the AD field never dominates the 
Universe in the above scenario because it decreases faster than that of 
radiation. 
Just after inflation, the AD field starts to oscillate around the origin 
due to the positive Hubble-induced mass term. 
Then, its number density decreases with time as $a^{-3}$ due to the expansion 
of the Universe. 
This means that its energy density decreases as $a^{-9/2}$ 
because its effective mass is of order the Hubble parameter, which decreases 
as $a^{-3/2}$. 
When the Hubble parameter decreases down to the mass of the AD field, 
that is, when $H(t) \simeq m_\phi$, 
its energy fraction to the total energy density is given as 
\beq
 \left. \frac{\rho_{\rm AD}}{\rho_{\rm tot}} \right\vert_{H \simeq m_\phi} 
 &\simeq& 
 \lmk \frac{m_\phi}{H_I} \rmk 
 \left. \frac{\rho_{\rm AD}}{\rho_{\rm tot}} \right\vert_{H \simeq H_I} 
 \nonumber \\ 
 & \simeq &
 10^{-11} \lmk \frac{\phi_i} {\Mpl} \rmk^2
 \lmk \frac{m_\phi}{\TEV} \rmk  
 \lmk \frac{H_I}{4\times 10^{12}\GEV} \rmk^{-1}.
\eeq
Thus, the energy density of AD field becomes negligible soon after inflation
and the result of Eq.~(\ref{Y_b}) is applicable to the case of 
$\phi_i \simeq \Mpl$.

\subsection{Baryonic isocurvature perturbations
\label{subsec:isocurvature}}

Although the initial phase and radial values of the flat direction 
are almost constant over the whole range of the observable universe, 
they acquire quantum fluctuations like~\cite{Enqvist:1998pf, Enqvist:1999hv, Kawasaki:2001in, Kasuya:2008xp, Harigaya:2014tla}
\beq
 \abs{\delta \theta_i} \simeq \frac{H_I }{2 \pi \abs{\phi_i}} \\
 \abs{\delta \phi_i} \simeq \frac{H_I }{2 \pi}. 
\eeq
These fluctuations result in baryonic isocurvature perturbations 
because the produced baryon density is related to the initial phase $\theta_i$ 
and VEV $\phi_i$ 
(see Eqs.~(\ref{n_B-L}) and (\ref{epsilon})). 
The baryonic isocurvature perturbation $S_{b\gamma}$ is given by 
\beq
 \mathcal{S}_{b \gamma} \equiv \frac{\delta Y_B}{Y_B} 
 \simeq n \lmk \cot \lmk n \theta \rmk \delta \theta + \frac{\delta \phi_i}{\phi_i} \rmk. 
 \label{S_b}
\eeq
Since we consider the case that the AD field starts to oscillate due to the 
positive Hubble-induced mass term, 
its radial direction also has quantum fluctuations and contributes 
to the baryonic isocurvature perturbations. 
This leads to an additional factor in Eq.~(\ref{S_b}), which cannot be 
suppressed by the tuning of the initial phase $\theta_i$. 
Note that 
the VEV of the AD field is smaller than the Planck scale 
due to the exponential factor in the supergravity potential. 
This leads to a lower bound on baryonic isocurvature perturbations like 
\beq
  \left\vert S_{{\rm b} \gamma} \right\vert 
  &\simeq& 
 2.7 \times 10^{-7} \times n \lmk \frac{H_I}{4 \times 10^{12} \GEV} \rmk \lmk \frac{\Mpl}{\phi_i} \rmk, 
 \nonumber \\   
  &\gtrsim& 
 2.7 \times 10^{-7} \times n \lmk \frac{H_I}{4 \times 10^{12} \GEV} \rmk, 
\eeq
where we assume $1 + \cot \lmk n \theta \rmk \approx 1$. 

Since the density perturbations of the cosmic microwave background 
are predominantly adiabatic~\cite{Hinshaw:2012aka, Ade:2013zuv}, 
the baryonic isocurvature perturbation is tightly constrained as~\cite{Ade:2013uln, Harigaya:2014tla} 
\beq
 \left\vert S_{{\rm b} \gamma} \right\vert \lesssim 
5.0 \times 10^{-5}. 
\eeq
Our scenario predicts the value below this constraint 
though it depends on the value of 
$n$ and $H_I$ (i.e., $\xi$). 
If one might consider a high-scale D-term inflation model, 
the resulting isocurvature perturbations can be as large as this lower bound 
and 
would be detected by future observations of CMB fluctuations.

\subsection{Comments on Q-ball formation
\label{subsec:Q-ball}}

In this subsection, we comment on Q-ball formation. 
If the potential of the AD field is shallower than the quadratic potential, 
its coherent oscillation is unstable and fragments into non-topological 
solitons, called Q-balls~\cite{Coleman}. 
The formation of Q-balls may change the scenario of Affleck-Dine 
baryogenesis~\cite{Qsusy, KuSh, EnMc, KK1, KK2, KK3, Higaki:2014eda}. 
For example, their decay can be another source of non-thermal production of 
DM~\cite{EnMc, Fujii:2001xp, Fujii:2002kr, Fujii:2002aj, 
Roszkowski:2006kw, Seto:2007ym, ShKu, Doddato:2011fz, Kasuya:2011ix, 
Doddato:2012ja, Kasuya:2012mh, KKY, Harigaya:2014tla, Kamada:2014ada}, 
or Q-balls can be a candidate for DM if they are 
stable~\cite{KuSh, Kasuya:2014ofa, Kasuya:2014bxa, Kasuya:2015uka}. 
In the case considered in this paper, 
the AD field starts to oscillate by the positive Hubble-induced mass term. 
When the beta function for the Hubble-induced mass of the AD field 
is positive, Q-ball does not form. 
The beta function has positive contributions from Yukawa interactions 
while it has negative ones from gauge interactions. 
The former positive contributions are roughly proportional to the squared 
masses of squarks and sleptons, 
and the latter negative ones are roughly proportional to the squared masses 
of gauginos. 
Here, since the Hubble-induced mass for gauginos is absent or 1-loop 
suppressed, 
the positive contributions from Yukawa interactions are usually dominant. 
Therefore, 
the beta function for the Hubble-induced mass of the AD field is usually 
positive 
and Q-balls may not form in our scenario. 
However, 
if the AD field consists only of the first and second family squarks 
and/or sleptons, 
the positive contributions from Yukawa interactions are suppressed by 
small Yukawa couplings. 
In this case, Q-balls might from. 
We estimate the typical charge $Q$ of Q-balls as 
\beq
 Q \sim \beta \lmk \frac{\phi_i}{m_{\phi,eff}} \rmk^2, 
\eeq
where $m_{\phi,eff}$ is the effective mass and 
$\beta$ ($\sim 10^{-2}$) is a numerical factor obtained from 
simulations of Q-ball formation~\cite{KK1, KK2, Hiramatsu:2010dx}. 
Here, we should substitute the Hubble-induced mass into the effective mass, 
and so the typical charge of Q-balls is at most $10^{8}$. 
Such small Q-balls soon evaporate into thermal plasma via interactions with the thermal plasma~\cite{Laine:1998rg, Banerjee:2000mb} 
(see also Ref.~\cite{KK3}).%
\footnote{
The evaporation is efficient during the inflaton oscillation era. 
In addition, 
since the energy per unit charge for these Q-balls is given by the Hubble parameter, 
their energy density decreases with time as $a^{-9/2}$. 
Thus, the energy density of the Q-balls never dominate that of the Universe. 
}
Therefore, the subsequent cosmological scenario and 
the calculation of the baryon asymmetry does not change.

Even if Q-balls do not form just after the end of inflation, 
they may form at the time of $H (t) \simeq m_\phi$. 
After that time, 
the potential of the AD field is dominated by its soft mass term. 
If the beta function of the soft mass is negative, 
the AD field becomes to fragment into Q-balls at that time. 
Since $n_b \propto H(t) \phi^2 (t) \propto a^{-3} \propto H(t)^2$ 
until the Hubble parameter decreases down to the soft mass, 
the amplitude of the AD field at $H(t) \simeq m_\phi$ is given as 
\beq
 \left. \phi \right\vert_{H(t) \simeq m_\phi} 
 \simeq \lmk \frac{\mphi}{H_I} \rmk^{1/2} \phi_i. 
\eeq
This implies that a typical charge of Q-balls is given as 
\beq
 Q \simeq \beta \lmk \frac{\phi}{m_\phi} \rmk^2
 \simeq \beta \lmk \frac{\phi_i^2}{m_\phi H_I} \rmk.
\eeq
This is at most $10^{18}$ for typical parameters. 
Such small Q-balls are evaporate into thermal plasma 
soon after they form. 
Even if Q-balls survive, they are so small as to decay into quarks 
before the BBN epoch. 
However, they usually decay after the electroweak phase transition~\cite{evap, KY}. 
Since the sphaleron process is decoupled at that time, 
the AD field has to carry a nonzero baryon charge (not $B-L$) to generate the baryon asymmetry. 
In that case, 
the resulting baryon asymmetry is given by Eq.~(\ref{Y_b}) 
without the factor of $8/23$.

\section{\label{sec:coincidence}
Model for solution to the baryon-DM coincidence problem}

In this section, 
we propose a D-term inflation model 
which predicts an $\mathcal{O}(1)$ ratio of baryon to DM density. 
We introduce a shift symmetry to ensure the flatness of the inflaton potential 
above the Planck scale. 
When there exists a small linear term in the \Kahler potential, 
the inflaton decays mainly into gravitinos~\cite{
Nakamura:2006uc, Kawasaki:2006gs, Asaka:2006bv, Dine:2006ii, Endo:2006tf, Kawasaki:2006hm, Endo:2006qk}. 
The subsequent decay of those gravitinos is a source of non-thermal production 
of LSP DM 
and the resulting DM abundance is proportional to the reheating temperature 
and inversely proportional to the inflaton mass. 
Since the amount of the baryon asymmetry in Eq.~(\ref{Y_b}) has similar 
parameter dependences, 
the baryon and DM densities are related with each other through the energy 
scale of inflation.

In the next subsection, we propose the D-term inflation model. 
In Sec.~\ref{subsec:reheating}, 
we investigate reheating processes of the D-term inflation model 
and then 
calculate the abundance of DM.

\subsection{Model
\label{subsec:model}}

Let us introduce a shift symmetry and an approximate $Z_2$ symmetry for the inflaton field $S$~\cite{Kawasaki:2000yn}. 
Under these symmetries, $S$ transforms as 
$S \rightarrow S + i\alpha$ ($\alpha$:real)
and $S \rightarrow -S$, respectively.
Then, the \Kahler potential is written as 
\beq
 K = 
 c_S \lmk S + S^* \rmk 
 + \half \lmk S + S^* \rmk^2 
 + \abs{\psi_-}^2 + \abs{\psi_+}^2, 
\eeq
where $c_S$ ($\ll 1$) is an order parameter for the $Z_2$ symmetry breaking 
effect. 
The superpotential of Eq.~(\ref{W_inf}) 
explicitly breaks the shift symmetry, 
which is required to ensure a graceful exit. 
Otherwise the inflaton stays at a certain VEV because it has a exactly flat potential. 
In this model, we should replace $\abs{S}^2$ with $(S + S^*)^2 /2$ 
for the calculations in Sec.~\ref{subsec:potential}, 
though the results are unchanged.

There is an advantage to impose the shift symmetry to the inflaton. 
In order to obtain a sufficiently large e-folding number, say, 
$N_* \gtrsim 60$, the initial VEV of the inflaton $S$ has to be as large as 
$N_* \frac{\sqrt{2} g^2}{4 \pi^2} \Mpl \simeq 0.5 \Mpl$. 
which is of order the Planck scale. 
This implies that the Planck-scale physics may affect 
the potential of the inflaton 
and 
spoil its flatness. 
However, the shift symmetry ensures the flatness of the inflaton potential 
above the Planck scale.

If $Z_2$ symmetry is exact,  
some MSSM particles have to carry odd $Z_2$ charge 
and interact with the field $S$~\cite{Kawasaki:2000yn} for the inflaton 
to decay. 
In this case 
the LSP DM abundance is given by the usual thermal relic density. 
Here, we introduce $Z_2$ breaking terms in the \Kahler potential 
so that the field $S$ efficiently decays into 
gravitinos~\cite{Kawasaki:2006hm, Endo:2006qk}, 
whose decay is a source of non-thermal production of LSP DM.

We assume that the mass of gravitinos is of order $10^{2-3} \TEV$ 
so that gravitinos decay into radiation before the BBN epoch. 
Otherwise the decay of gravitinos spoils the success of the BBN, 
or their energy density overcloses the Universe if they are stable. 
Such a heavy gravitino is well motivated 
in a class of SUSY models with a split spectrum~\cite{ArkaniHamed:2004fb, Giudice:2004tc, 
ArkaniHamed:2004yi, Wells:2004di, Hall:2011jd, Ibe:2011aa, Ibe:2012hu}. 
In these models, 
the masses of gravitino as well as squarks and sleptons are of order 
(or larger than) $10^{2-3} \TEV$ 
while those of gauginos are of order $1\TEV$. 
This hierarchy can be realized 
when gauginos acquire one-loop suppressed soft masses through the anomaly 
mediated SUSY breaking effect~\cite{Randall:1998uk, Giudice:1998xp}. 
In that case, the mass of wino and gravitino is related with each other 
such as 
\beq 
 m_{\tilde{w}} = \frac{g_2^2}{16 \pi^2} \lmk \mg + L \rmk 
 \simeq 3 \times 10^{-3} \lmk \mg + L \rmk. 
\eeq
The factor $L$ is the Higgsino threshold corrections and is calculated as~\cite{Giudice:1998xp, Gherghetta:1999sw}
\beq
 L \equiv \mu_H \sin 2 \beta \frac{m_A^2}{\abs{\mu_H}^2 - m_A^2 } \log \frac{\abs{\mu_H}^2}{m_A^2}, 
\eeq
where $m_A$ is the mass of the heavy Higgs bosons, 
$\mu_H$ is the SUSY mass of the higgsinos, 
and $\tan \beta$ is the ratio of the VEV of $H_u$ and $H_d$. 
When $\mu_H$ is of order the gravitino mass, 
the Higgsino threshold correction is important and 
the wino mass is $\sim 10^{-3} \mg$. 
Note that neutral higgsino can also be the LSP when $\mu_H$ is sufficiently 
small. 
The following discussion does not rely on the detailed properties of the LSP 
except for its mass. 
Hereafter, we assume that the mass of gravitino is $O(10^{2-3}) \TEV$ and 
that of the LSP is $O(10^{2-3}) \GEV$.

In order to calculate the gravitino production rate from the inflaton decay, 
we need to specify the SUSY breaking sector. 
We introduce a Polonyi field $z$, which 
breaks SUSY in low energy scale, 
and consider a simple extension of the Polonyi model given as 
\beq
 K = \abs{z}^2 - \frac{\abs{z}^4}{\Lambda^2}, \\
 W = \mu^2 z + W_0, 
\eeq
where $\Lambda$ is a cutoff scale, $\mu$ is the SUSY breaking scale, 
and $W_0$ is a constant term which makes the cosmological constant (almost) 
zero in the present Universe. 
This can be achieved by the O'Raifeartaigh model after integrating out 
relatively heavy particles~\cite{Kallosh:2006dv}
or by dynamical SUSY breaking models, including the IYIT model~\cite{Izawa:1996pk, Intriligator:1996pu}. 
The important parameters are calculated as 
\beq
 \mu^2 &\simeq& \sqrt{3} \mg \Mpl, \\
 m_z^2 &\simeq& \frac{12 \mg^2}{\Lambda^2} \Mpl^2, \label{m_z}\\
 \la z \ra_0 &\simeq& 2 \sqrt{3} \lmk \frac{\mg}{m_z} \rmk^2 \Mpl, \label{z VEV} 
\eeq
where $m_z$ is the mass of $z$ and $\la z \ra$ is its VEV at the low energy 
vacuum. 
Since the Hubble-induced mass is absent during D-term inflation, 
massless scalar fields cannot be stabilized at the origin. 
This implies that 
if the mass of the Polonyi field is much smaller than the Hubble parameter, 
it obtains a VEV as large as the Planck scale during inflation. 
In this section, we consider the case that 
the mass of the Polonyi field is larger than $H_I$ 
and it stays at the origin of the potential during inflation, 
while we consider the case of relatively light Polonyi in Appendix. 
The conditions of $m_z \gtrsim H_I$ and $\Lambda \gtrsim \mu$ in the effective 
theory leads to the lower bound on the gravitino 
mass~\cite{Buchmuller:2013uta}: 
\beq
 \mg &\gtrsim& \frac{\sqrt{3} H_I^2}{12 \Mpl}, 
 \nonumber \\
 &\simeq& 10^3 \TEV \lmk \frac{H_I}{4 \times 10^{12} \GEV} \rmk^2. 
\eeq
Thus, the heavy gravitino is favoured in D-term inflation to stabilize the VEV of the Polonyi field.

\subsection{Reheating process
\label{subsec:reheating}}

In this subsection, 
we investigate the reheating process and 
calculate the DM abundance in the model introduced in the previous subsection.

As explained in Sec.~\ref{sec:D-term inflation}, 
the field $\psi_+$ decays into the MSSM fields much faster than the inflaton 
$S$, 
so that 
the reheating temperature of the Universe is determined by the relatively 
late-time decay of the inflaton $S$~\cite{Kolda:1998kc}. 
After the field $\psi_+$ decays completely, 
the effective superpotential can be rewritten as 
\beq
 W^{(\rm inf)} &=& m_S S \psi_-, \\
 m_S &\equiv& \lambda \sqrt{\xi}, 
\eeq
where $m_S$ is the effective mass of the fields $S$ and $\psi_-$. 
This superpotential is equivalent to the one in the model of chaotic inflation 
proposed in Ref.~\cite{Kawasaki:2000yn}, 
except for the value of $m_S$.

The supergravity effects induce a soft SUSY breaking B-term of 
$b m_{3/2} m_S S \psi_-$, 
where $b$ is an $\mathcal{O}(1)$ constant. 
This implies that 
they maximally mix with each other 
and form mass eigenstates 
\beq
 \Phi_\pm \equiv \frac{1}{\sqrt{2}} \lmk S \pm \psi_-^\dagger \rmk, 
 \label{Phi_pm} 
\eeq
around the potential minimum~\cite{Kawasaki:2006gs, Kawasaki:2006hm}. 
Therefore, 
when the time scale of inflaton decay $\Gamma_S^{-1}$ 
is longer than that of the mixing effect $\mg^{-1}$, 
we have to consider the decay of $\Phi_\pm$ to investigate the reheating 
process. 
Since we consider a heavy gravitino and a reheating temperature of order 
$10^3 \GEV$ (see Eq.~(\ref{Y_b})), 
the mixing effect is indeed relevant.

The $Z_2$ breaking term in the \Kahler potential 
results in the decay of 
the field $\Phi_\pm$ through supergravity effects~\cite{Endo:2006qk}.%
\footnote{
Since the field $\psi_-$ has a small VEV of order $c \mg / m_S$ 
due to the supergravity effects, 
it can decay into the MSSM fields through the D-term potential. 
However, 
we confirm that 
its partial decay rate is irrelevant due to the suppression factor coming 
from its small VEV and can be neglected. 
}
First, let us focus on the top Yukawa interaction in the MSSM sector: 
\beq
 W^{(\rm top)} = y_t Q_3 H_u u^c_3, 
\eeq
where $y_t$ is the top Yukawa coupling constant, 
and $Q_3$, $H_u$, and $u^c_3$ are the chiral supermultiplets of the MSSM sector. 
The relevant interaction terms between 
$\psi_-$ and the MSSM fields are given by 
\beq
 V &=& 
\frac{1}{\Mpl}
 K_S W^{(\rm top)} 
 W_{S}^* + c.c. + \dots, 
 \nonumber \\ 
 &=& \frac{y_t m_S K_S}{\Mpl^2}
 \psi_-^* (\tilde{Q}_3 H_u \tilde{u}^c_3) + c.c. + \dots, 
 \label{V interaction}
\eeq
where 
the dots "$\dots$" represents the other irrelevant terms. 
Since the fields $\Phi_\pm$ consist of $\psi_-$ as Eq.~(\ref{Phi_pm}), 
they decay into the MSSM scalar fields through this interaction. 
They also decay into the MSSM fermion fields, 
which equally contributes to the $\Phi_\pm$ decay~\cite{Endo:2006qk}. 
Thus, the partial decay rate of $\Phi_\pm$ into the MSSM fields is given as 
\beq
 \Gamma_{\rm MSSM} \lmk \Phi_\pm \to {\rm MSSM} \rmk = \frac{3 c_S^2}{256 \pi^3} \abs{y_t}^2 
 \frac{m_S^3}{\Mpl^2}, 
 \label{gamma_MSSM}
\eeq
where we use $K_S = c_S$. 
Since we consider the gaugino mass ($m_{\tilde{g}}$) much smaller than the gravitino mass, 
the decay rates of $S$ into gauge fields are suppressed by a factor of $(m_{\tilde{g}} / \mg)^2$ 
and can be neglected~\cite{Endo:2006tf}.

Next, let us consider the decay of $\Phi_\pm$ into gravitinos~\cite{
Nakamura:2006uc, Kawasaki:2006gs, Asaka:2006bv, Dine:2006ii, Endo:2006tf, Kawasaki:2006hm, Endo:2006qk}. 
We follow the discussion presented in Ref.~\cite{Nakayama:2012hy}. 
When the field $S$ has a nonzero VEV, 
the field $\psi_-$ mixes with the SUSY breaking field $z$ 
and can decays into goldstino, 
(i.e., longitudinal component of gravitino). 
This is because the supergravity effects induce mixing terms such as 
\beq
 V &=& W_S  \lmk K_S W \rmk^* 
 + K_{S \bar{z}}^{-1} W_S W_z^* 
 + c.c. + \dots 
 \nonumber \\
 &=& m_S d F_z  \psi_- z^* + c.c. + \dots, \\
 d &\equiv& \la K_S \ra - \la K_{S z \bar{z}} \ra, 
\eeq
where the dots "$\dots$" represent the other irrelevant terms. 
The second term in $d$ is relevant when 
there is a term like $(S + S^*) \abs{z}^2$ in the \Kahler potential, 
whose coefficient is of order $c_S$. 
Thus, 
the fields $\psi_-$ and $z$ mix with each other 
and the mixing angle is given by 
\beq
 \theta \simeq d \frac{F_z m_S}{m_z^2}, 
\eeq
where we use $m_z \gg m_S$. 
Since the fields $\Phi_\pm$ consist of $\psi_-$, 
they mix with $z$ 
and the mixing angle is given by $\theta / \sqrt{2}$. 
Since the SUSY breaking field $z$ has an operator of 
\beq
 \mathcal{L} 
 = - 2 \frac{F_z}{\Lambda^2} z \tilde{z}^\dagger \tilde{z}^\dagger + h.c., 
\eeq
it decays into goldstino $\tilde{z}$. 
Together with the mixings 
between $\Phi_\pm$ and $z$, 
the fields $\Phi_\pm$ decays into goldstino through this operator. 
The partial decay rate of the field $\Phi_\pm$ into goldstino 
is therefore calculated as~\cite{Nakayama:2014xca}
\beq
 \Gamma_{\tilde{z}} \lmk \Phi_\pm \to \tilde{z} \tilde{z} \rmk &\simeq& 
 \frac{1}{32 \pi} \lmk \frac{\theta}{\sqrt{2}} \rmk^2 \frac{m_z^4}{\abs{F_z}^2} m_S, \nonumber \\ 
&\simeq& 
 \frac{d^2}{64 \pi} \frac{m_S^3}{\Mpl^2}. 
 \label{gamma_3/2} 
\eeq

From Eqs. (\ref{gamma_MSSM}) and (\ref{gamma_3/2}), 
the total decay rate $\Gamma_S$ and 
the branching ratio of the decay of $\Phi_\pm$ into gravitinos 
$B_{3/2}$ are given by
\beq
 \Gamma_S &=& \Gamma_{\rm MSSM} \lmk \Phi_\pm \to {\rm MSSM} \rmk 
 + \Gamma_{\tilde{z}} \lmk \Phi_\pm \to \tilde{z} \tilde{z} \rmk, \\ 
 {\rm Br}_{3/2} &=& \frac{d^2}{d^2 + 3 \abs{y_t}^2 c^2/(4 \pi^2)}. 
 \label{B_3/2}
\eeq
Since $d/c = \mathcal{O}(1)$ and $y_t = \mathcal{O}(1)$, 
the branching ratio is almost unity. 
This means that 
the energy density of the Universe is dominated by that of the gravitinos 
after the fields $\Phi_\pm$ (i.e., the inflaton $S$) decay completely.%
\footnote{
Note that 
since we consider relatively low reheating temperature 
$\sim 10^{3-4} \GEV$, 
we can neglect the thermal production of 
gravitinos~\cite{Bolz:2000fu, Pradler:2006qh}. 
}

Since the fields $\Phi_\pm$ are much heavier than gravitino, 
the produced gravitinos are highly relativistic. 
The Lorentz factor for the gravitinos at a time $H^{-1} (t)$ is given as 
\beq
 \gamma (t) &=& 
 \lkk \lmk \frac{m_S}{\mg} \rmk^2 \frac{H(t)}{\Gamma_S} + 1 \rkk^{1/2} 
 \simeq \frac{m_S}{\mg} \lmk \frac{H(t)}{\Gamma_S} \rmk^{1/2}.
\eeq
The gravitinos decay into MSSM particles 
with a rate of 
\beq
 \Gamma_{3/2} \simeq \gamma^{-1} (t) \frac{1}{48 \pi} \frac{\sum_{i} m_{\tilde{X}_i}^5}{\mg^2 \Mpl^2}, 
\eeq
where the summation is taken for all MSSM particles $\tilde{X}_i$. 
Since we consider a SUSY model with relatively light gauginos and relatively heavy squark and sleptons, 
we can roughly estimate the numerator as $24 \mg^5$. 
This implies that the gravitino decays into radiation 
at the temperature
\beq
 T_{3/2} 
 &\simeq& \lmk \frac{90}{g_* \pi^2} \rmk^{1/4} \sqrt{\Gamma_{3/2} \Mpl} 
 \nonumber \\
 &\simeq& 
 1.1 \MEV 
 \lmk \frac{ T_{\rm RH} } {4 \times 10^3 \GEV} \rmk^{1/3} 
 \lmk \frac{ m_S } {5 \times 10^{15} \GEV} \rmk^{-1/3} 
 \lmk \frac{ m_{3/2} } {400 \TEV} \rmk^{4/3}, 
 \label{T_3/2}
\eeq
where $g_*$ ($\simeq 10.75$) is the effective number of degrees of freedom 
at the decay time. 
We require that the mass of gravitino is of order $10^{2-3} \TEV$ or larger 
so that its decay completes before the BBN epoch, that is, 
$T_{3/2} \gtrsim 1 \MEV$. 
Otherwise the decay particles interact with the light elements 
and spoil the success of the BBN~\cite{Kawasaki:1999na, Kawasaki:2000en, 
Kawasaki:2004qu, Ichikawa:2005vw, Kawasaki:2008qe}. 
The gravitino decay temperature $T_{3/2}$ 
is much smaller than the mass of the LSP, 
so that the decay of gravitino is a source of its nonthermal production. 
Since the energy density of the Universe is dominated by that of gravitino 
before they decay, 
the thermal relic density of the LSP is diluted by the entropy production from 
the gravitino decay. 
Therefore, the LSP abundance is determined by the nonthermal production 
from the gravitino decay. 
The produced number density of the LSPs is equal to that of the gravitinos
due to the R-parity conservation. 
Note that the annihilation of the produced LSP is usually inefficient 
in such a low temperature.

The Lorentz factor of the gravitino is 
of order $10^3$ for the reference parameters shown in Eq.~(\ref{T_3/2}). 
This implies that the scale factor of the Universe continues to decrease as 
$a^{-4}$ 
from the time of reheating by the decay of $\Phi_\pm$. 
Although the LSPs are relativistic at the time they are produced from 
gravitino decay, 
they lose their energy through interactions with the thermal plasma 
and soon become to non-relativistic 
particles~\cite{Hisano:2000dz, Arcadi:2011ev, Ibe:2012hr}. 
Therefore, the LSP DM is cold even though they are produced non-thermally 
in this scenario.

\subsection{DM density and baryon-DM coincidence
\label{subsec:DM}}

Let us summarize the scenario of non-thermal production of DM. 
First, the inflaton $S$ (or $\Phi_\pm$) decays into gravitinos as well as 
the MSSM particles at $H (t) \simeq \Gamma_S$. 
Then the energy density of the Universe is dominated by the relativistic 
gravitinos and decreases as $a^{-4}$. 
The gravitinos decay into the MSSM particles just before the epoch of the BBN 
and the LSP DM is produced non-thermally. 
Since the thermal relic density of the LSP is diluted by the entropy 
production of gravitino decay, 
its abundance is determined by the gravitino decay. 
Thus, we can estimate the resultant DM abundance as 
\beq
 Y_{\rm DM} 
 &\equiv& \frac{n_{\rm LSP}}{s} \nonumber \\
 &\simeq& \left. \frac{n_{3/2}}{s} \right\vert_{H=\Gamma_{3/2}} 
 \nonumber \\
 &\simeq& \left. \frac{3 T_{3/2}}{4} 
  \frac{n_{3/2}}{\rho_{3/2}} \right\vert_{H=\Gamma_{3/2}} 
  \nonumber \\
 &\simeq& \left. \frac{3 T_{3/2}}{4} 
  \lmk \frac { \Gamma_S } { \Gamma_{3/2} } \rmk^{1/2} 
  \frac{ n_{3/2}}{\rho_{3/2} } \right\vert_{H=\Gamma_{S}} 
  \nonumber \\
 &\simeq& \left. 
 \frac{3 T^{\rm (eff)}_{\rm RH}}{4}
 \frac{2 {\rm Br}_{3/2} n_S}{\rho_S} \right\vert_{H=\Gamma_{S}} 
 \nonumber \\ 
 &\simeq& \frac{3 T^{\rm (eff)}_{\rm RH} }{2 m_S}, 
\label{Y_DM}
\eeq
where we have used ${\rm Br}_{3/2} \simeq 1$ in the last line. 
We define the effective reheating temperature $T^{\rm (eff)}_{\rm RH}$ 
by Eq.~(\ref{T_RH}) 
with the replacement of $g_* (T_{\rm RH}) \to g_* (T_{3/2})$ as
\beq
 T^{\rm (eff)}_{\rm RH} 
 &\simeq& \lmk \frac{90}{g_*(T_{3/2}) \pi^2} \rmk^{1/4} \sqrt{\Gamma_S \Mpl} 
 \nonumber \\
 &\simeq& 
 1.5 \times 10^{3} \GEV 
 \lmk \frac{m_S}{5 \times 10^{15} \GEV} \rmk^{3/2} 
 \lmk \frac{d}{10^{-10}} \rmk. 
 \label{effective T_RH}
\eeq
The reheating temperature is adjusted by the $Z_2$ symmetry order parameter 
$d$ to obtain a desirable amount of baryon asymmetry from Eq.~(\ref{Y_b}) 
or DM from Eq.~(\ref{Y_DM}).

Here we take into account the baryon asymmetry generated by 
the Affleck-Dine baryogenesis. 
Once we replace the reheating temperature $T_{\rm RH}$ with the effective one 
$T^{\rm (eff)}_{\rm RH}$ defined by Eq.~(\ref{effective T_RH}), 
the resulting baryon asymmetry is still given by Eq.~(\ref{Y_b}) 
even in this scenario.%
\footnote{
Note that the inflaton also decays into the MSSM fields with a branching 
$\sim 10^{-(1-2)}$, 
so that there exists significant thermal plasma after the decay of inflaton. 
Thus, the sphaleron effect proceeds fast enough to convert the $B-L$ 
asymmetry to the baryon asymmetry 
even if the dominant component of the Universe is gravitino at that time. 
}
Combining Eqs.~(\ref{Y_b}) and (\ref{Y_DM}), 
we obtain the following simple relation for the baryon-to-DM ratio: 
\beq
 \frac{\Omega_b}{\Omega_{\rm DM}} 
 \simeq 
 \frac{4}{69}
 \epsilon q
 \frac{m_p}{m_{\rm LSP}} 
 \frac{m_S}{H_I}, 
\eeq
where we assume $\phi_i \simeq \Mpl$. 
Substituting benchmark parameters and the proton mass $m_p \simeq 0.938 \GEV$, 
we obtain 
\beq
 \frac{\Omega_b}{\Omega_{\rm DM}} 
 &\simeq& 
 0.22 
 \epsilon q 
 \lmk \frac{m_{\rm LSP}} { 400 \GEV } \rmk^{ -1 }
 \lmk \frac{ m_S } { 6.6 \times 10^{15} \GEV } \rmk
 \lmk \frac{ H_I } { 4 \times 10^{12} \GEV } \rmk^{-1}, 
 \nonumber \\
 &\simeq& 0.12 
 \epsilon q \lambda g^{-1} 
 \lmk \frac{m_{\rm LSP}} { 400 \GEV } \rmk^{ -1 }
 \lmk \frac{ \sqrt{\xi} } { 6.6 \times 10^{15} \GEV } \rmk^{-1}, 
 \label{b-DM ratio}
\eeq
which is naturally of order unity 
and is consistent with the observed value of 
$\Omega_b^{(\rm obs)} / \Omega_{\rm DM}^{(\rm obs)} \simeq 0.2$~\cite{pdg}. 
The scenario naturally explains 
the coincidence of their energy density, known as the baryon-DM coincidence 
problem. 
This is because both of them are related to the energy scale of inflation. 
The amount of baryon asymmetry is proportional to the reheating temperature 
of the Universe 
and inversely proportional to the Hubble parameter during inflation. 
That of DM is proportional to the reheating temperature 
and inversely proportional to the mass of inflaton. 
Since the Hubble parameter ($H_I \sim g\xi/\Mpl$) and 
the inflaton mass ($m_S = \lambda\sqrt{\xi}$) is related to each other, 
the resulting baryon and DM density is naturally of order unity. 
Interestingly, 
the electroweak scale DM mass ($O(10^2) \GEV$) comes from the fact that 
the GUT scale, which $\sqrt{\xi}$ is expected to be, is two orders of magnitude less than the Planck scale.

Although the result has an $\mathcal{O}(1)$ uncertainty coming mainly from 
$\lambda$ and $\xi$, 
the LSP with mass of $O(10^{2-3}) \GEV$ is favoured in our scenario. 
If the LSP DM is mostly wino or higgsino, 
the indirect detection experiments of DM puts lower bounds 
on DM mass. 
The wino DM with $m_{\tilde{w}} \le 390 \GEV$ and 
$2.14 \TEV \le m_{\tilde{w}} \le 2.53 \TEV$ 
is excluded~\cite{Bhattacherjee:2014dya}, while 
the higgsino DM with $m_{\tilde{h}} \le 160 \GEV$ is 
excluded~\cite{Fan:2013faa}. 
The future indirect detection experiments 
can detect the wino DM with $m_{\tilde{w}} \le 1.0 \TEV$ and 
$1.66 \TEV \le m_{\tilde{w}} \le 2.77 \TEV$~\cite{Bhattacherjee:2014dya}.

\section{Summary
\label{sub:summary}}

We have considered the Affleck-Dine baryogenesis 
assuming a positive Hubble-induced mass term after D-term inflation. 
We have calculated the baryon asymmetry 
and found that 
the result is consistent with the observed abundance of baryon asymmetry 
for reheating temperature $T_{\rm RH} \sim 10^{3} \GEV$. 
There are some differences from the conventional scenarios of the Affleck-Dine 
baryogenesis 
where the sign of the Hubble-induced mass term is negative. 
First, since the AD field starts to oscillate just after the end of inflation, 
the resulting baryon asymmetry is inversely proportional to the inflation 
scale $H_I$. 
This allows us to consider a relatively large reheating temperature 
even if the initial VEV $\phi_i$ is as large as the Planck scale. 
In addition, the ellipticity parameter $\epsilon$, which describes the 
efficiency of baryogenesis, 
can be much smaller than unity when $\phi_i$ is smaller than the Planck scale. 
This is because the phase direction of the AD field is kicked by a 
higher-dimensional \Kahler potential, 
which effect is highly suppressed for a small VEV of the AD field. 
Since the radial direction as well as the phase one has quantum fluctuations 
during D-term inflation, 
the Affleck-Dine baryogenesis predicts nonzero baryonic isocurvature 
perturbations.
They would be detected by future CMB observations 
if one considers a high-scale D-term inflation model.

We also proposed a D-term inflation model 
with a shift symmetry in the imaginary direction of the inflaton superfield
and a small linear term in the \Kahler potential. 
We consider the case that the mass of gravitino is $O(10^{2-3}) \TEV$ 
and the mass of the LSP is $O(10^{2-3}) \GEV$, 
which is naturally realized in anomaly mediated SUSY breaking models. 
In this model, the inflaton decays mainly into gravitinos through the supergravity effects~\cite{
Nakamura:2006uc, Kawasaki:2006gs, Asaka:2006bv, Dine:2006ii, Endo:2006tf, Kawasaki:2006hm, Endo:2006qk} 
and the subsequent decay of gravitinos is a source of non-thermal 
production of DM. 
Together with estimation of baryon asymmetry generated from 
the Affleck-Dine mechanism, 
the resulting DM density gives an $\mathcal{O}(1)$ baryon-to-DM ratio. 
This is because both of them are related to the energy scale of inflation. 
The amount of baryon asymmetry is proportional to the reheating temperature 
of the Universe 
and inversely proportional to the Hubble parameter during inflation. 
That of DM is proportional to the reheating temperature 
and inversely proportional to the mass of inflaton. 
Since the Hubble parameter and the mass is related to each other, 
the resulting baryon and DM density is naturally of order unity. 
We predict that the LSP mass is 
two orders of magnitude larger than the proton mass, 
which comes from the fact that 
the GUT scale is two orders of magnitude less than the Planck scale. 
When the LSP is mostly wino or higgsino, 
it would be detected by future indirect detection experiments of DM.

\vspace{1cm}

\section*{Acknowledgments}
This work is supported by Grant-in-Aid for Scientific research 
from the Ministry of Education, Science, Sports, and Culture
(MEXT), Japan, No. 25400248 (M.K.), 
World Premier International Research Center Initiative
(WPI Initiative), MEXT, Japan,
and the Program for the Leading Graduate Schools, MEXT, Japan (M.Y.).
M.Y. acknowledges the support by JSPS Research Fellowship for Young Scientists.

\vspace{1cm}

\appendix


\section{Another solution for baryon-DM coincidence problem
\label{appendix}} 

In this appendix, we consider the case that 
the mass of the Polonyi field is less than the Hubble parameter $H_I$ 
and it obtains a nonzero VEV during inflation. 
Even if the origin of its potential is a symmetry-enhanced point, 
the Polonyi field cannot be stabilized at the origin 
because of the absence of the Hubble-induced mass during D-term inflation. 
Although 
one might wonder if this is an obstacle known as the Polonyi problem, 
we show that its decay 
can explain the amount of DM 
once we allow an $10\%$ fine-tuning for the initial VEV of the Polonyi field.

During D-term inflation, 
the Polonyi field obtains a VEV denoted as $z_i$, 
which might be as large as the Planck scale. 
Just after inflation ends, the Polonyi field starts to oscillate around the origin 
due to the positive Hubble-induced mass term as the AD field considered in Sec.~\ref{sec:ADBG}. 
Then, its number density decreases with time as $a^{-3}$ due to the expansion of the Universe. 
This means that its energy density decreases as $a^{-9/2}$ 
because its Hubble-induced mass is of order the Hubble parameter, which decreases as $a^{-3/2}$. 
After the Hubble parameter decreases down to its low-energy mass of the Polonyi field, 
that is, after the time of $H(t) \simeq m_z$, 
the Hubble-induced mass term can be neglected. 
Then its mass and VEV are given by Eqs~(\ref{m_z}) and (\ref{z VEV}), respectively. 
Note that the minimum of the potential is usually much smaller than 
the amplitude of the Polonyi field at that time 
because its amplitude decreases only as $a^{-3/4}$ until $H(t) \simeq m_z$: 
\beq
 \left. z \right\vert_{H (t) = m_z} \simeq \lmk \frac{m_z}{H_I} \rmk^{1/2} z_i 
 \gg \la z \ra_0. 
\eeq
This means that its number density is not affected 
and continues to decreases with time as $a^{-3}$.

The Polonyi field decays mainly into gravitinos with a rate of 
\beq
 \Gamma_z (z \to 2 \psi_{3/2} ) \simeq \frac{1}{96 \pi} \frac{m_z^5}{\mg^2 \Mpl^2}. 
\eeq
Since we consider relatively low reheating temperature to realize the Affleck-Dine baryogenesis, 
we can neglect the thermal production of gravitinos. 
Thus, the gravitino abundance is determined by the number density of the Polonyi field. 
We require that the mass of gravitino is of order $10^{2-3} \TEV$ 
so that it decays into the MSSM particles before the BBN epoch. 
Otherwise the decay particles interact with the light elements 
and spoil the success of the BBN~\cite{Kawasaki:1999na, Kawasaki:2000en, 
Kawasaki:2004qu, Ichikawa:2005vw, Kawasaki:2008qe}. 
The decay of gravitino is a source of non-thermal production of LSP DM. 
We assume that the thermal relic of the LSP is much smaller than 
the observed DM abundance. 
This can be achieved for the case of wino-like LSP with a mass much less than $3 \TEV$~\cite{Hisano:2006nn, Cirelli:2007xd} 
or higgsino-like LSP with a mass much less than $1 \TEV$. 
The wino-like LSP is well motivated in models of anomaly mediated SUSY breaking, 
where the gravitino mass is naturally as large as $O(100) \TEV$ as we required.

In summary, 
the LSP DM is non-thermally produced from the decay of gravitino, 
which is generated from the decay of the Polonyi field. 
The Polonyi field is generated by its coherent oscillation just after the end of inflation. 
We thus obtain the following DM abundance: 
\beq
 Y_{\rm DM} 
 &\equiv& \frac{n_{\rm LSP}}{s} 
 \nonumber \\
 &\simeq& \left. \frac{n_{3/2}}{s} \right\vert_{\psi_{3/2} {\rm decay}} 
 \nonumber \\
 &\simeq& \left. \frac{3 T_{\rm RH} n_{3/2}}{4 \rho_I} \right\vert_{\rm RH} 
 \nonumber \\
 &\simeq& \left. \frac{3 T_{\rm RH} n_z}{2 \rho_I} \right\vert_{H = \Gamma_z} 
 \nonumber \\
 &\simeq& \left. \frac{3 T_{\rm RH} n_z}{2 \rho_I} \right\vert_{H= H_I} 
 \nonumber \\
 &\simeq& \frac{T_{\rm RH} z_i^2}{2 H_I \Mpl^2}, 
\eeq
where we use $n_z \simeq H_I z_i^2$ at $H = H_I$ in the last line. 

Together with estimation of baryon asymmetry generated by the Affleck-Dine mechanism 
calculated in Sec.~\ref{subsec:ADBG}, 
the above result implies that 
the baryon-to-DM ratio is given by the following simple relation: 
\beq
 \frac{\Omega_b}{\Omega_{\rm DM}} 
 \simeq 
 \frac{4}{23} 
 \epsilon b 
 \frac{m_p}{m_{\rm LSP}} 
 \lmk \frac{\phi_i}{z_i} \rmk^{2}. 
\eeq
Thus, we can explain the observed baryon-to-DM ratio 
when $z_i/\phi_i \sim 0.1$. 
One might expect that the natural values of their initial VEVs are of order the Planck scale. 
In that case, the result requires a $10\%$ fine-tuning for the value of $z_i$.




\begin{thebibliography}{90}

  
  
\bibitem{pdg} 
  J.~Beringer {\it et al.}  [Particle Data Group Collaboration],
  Phys.\ Rev.\ D {\bf 86}, 010001 (2012).
  


\bibitem{Binetruy:1996xj} 
  P.~Binetruy and G.~R.~Dvali,
  Phys.\ Lett.\ B {\bf 388}, 241 (1996)
  [hep-ph/9606342].
  
\bibitem{Halyo:1996pp} 
  E.~Halyo,
  Phys.\ Lett.\ B {\bf 387}, 43 (1996)
  [hep-ph/9606423].



\bibitem{Hinshaw:2012aka} 
  G.~Hinshaw {\it et al.}  [WMAP Collaboration],
  Astrophys.\ J.\ Suppl.\  {\bf 208}, 19 (2013)
  [arXiv:1212.5226 [astro-ph.CO]].
    
  
\bibitem{Ade:2013zuv} 
  P.~A.~R.~Ade {\it et al.}  [Planck Collaboration],
  arXiv:1303.5076 [astro-ph.CO].


\bibitem{AD}
  I.~Affleck and M.~Dine,
  Nucl.\ Phys.\  B {\bf 249}, 361 (1985).

\bibitem{DRT} 
  M.~Dine, L.~Randall and S.~D.~Thomas,
  Nucl.\ Phys.\ B {\bf 458}, 291 (1996).
  [hep-ph/9507453].




\bibitem{Kolda:1998kc} 
  C.~F.~Kolda and J.~March-Russell,
  Phys.\ Rev.\ D {\bf 60}, 023504 (1999)
  [hep-ph/9802358].

\bibitem{Enqvist:1998pf} 
  K.~Enqvist and J.~McDonald,
  Phys.\ Rev.\ Lett.\  {\bf 83}, 2510 (1999)
  [hep-ph/9811412].

\bibitem{Enqvist:1999hv} 
  K.~Enqvist and J.~McDonald,
  Phys.\ Rev.\ D {\bf 62}, 043502 (2000)
  [hep-ph/9912478].

\bibitem{Kawasaki:2001in} 
  M.~Kawasaki and F.~Takahashi,
  Phys.\ Lett.\ B {\bf 516}, 388 (2001)
  [hep-ph/0105134].



\bibitem{Kuzmin:1985mm} 
  V.~A.~Kuzmin, V.~A.~Rubakov and M.~E.~Shaposhnikov,
  Phys.\ Lett.\ B {\bf 155}, 36 (1985).
  
  
\bibitem{Fukugita:1986hr} 
  M.~Fukugita and T.~Yanagida,
  Phys.\ Lett.\ B {\bf 174}, 45 (1986).
  
  
\bibitem{Kasuya:2008xp} 
  S.~Kasuya, M.~Kawasaki and F.~Takahashi,
  JCAP {\bf 0810}, 017 (2008)
  [arXiv:0805.4245 [hep-ph]].

\bibitem{Harigaya:2014tla} 
  K.~Harigaya, A.~Kamada, M.~Kawasaki, K.~Mukaida and M.~Yamada,
  Phys.\ Rev.\ D {\bf 90}, no. 4, 043510 (2014)
  [arXiv:1404.3138 [hep-ph]].
  
  
\bibitem{Ade:2013uln} 
  P.~A.~R.~Ade {\it et al.}  [Planck Collaboration],
  arXiv:1303.5082 [astro-ph.CO].


\bibitem{Kawasaki:2000yn} 
  M.~Kawasaki, M.~Yamaguchi and T.~Yanagida,
  Phys.\ Rev.\ Lett.\  {\bf 85}, 3572 (2000)
  [hep-ph/0004243].
  
  

\bibitem{Nakamura:2006uc} 
  S.~Nakamura and M.~Yamaguchi,
  Phys.\ Lett.\ B {\bf 638}, 389 (2006)
  [hep-ph/0602081].

\bibitem{Kawasaki:2006gs} 
  M.~Kawasaki, F.~Takahashi and T.~T.~Yanagida,
  Phys.\ Lett.\ B {\bf 638}, 8 (2006)
  [hep-ph/0603265].
  
\bibitem{Asaka:2006bv} 
  T.~Asaka, S.~Nakamura and M.~Yamaguchi,
  Phys.\ Rev.\ D {\bf 74}, 023520 (2006)
  [hep-ph/0604132].
  
  
\bibitem{Dine:2006ii} 
  M.~Dine, R.~Kitano, A.~Morisse and Y.~Shirman,
  Phys.\ Rev.\ D {\bf 73}, 123518 (2006)
  [hep-ph/0604140].
  
\bibitem{Endo:2006tf} 
  M.~Endo, K.~Hamaguchi and F.~Takahashi,
  Phys.\ Rev.\ D {\bf 74}, 023531 (2006)
  [hep-ph/0605091].
  
  
\bibitem{Kawasaki:2006hm} 
  M.~Kawasaki, F.~Takahashi and T.~T.~Yanagida,
  Phys.\ Rev.\ D {\bf 74}, 043519 (2006)
  [hep-ph/0605297].
  
  
  
\bibitem{Endo:2006qk} 
  M.~Endo, M.~Kawasaki, F.~Takahashi and T.~T.~Yanagida,
  Phys.\ Lett.\ B {\bf 642}, 518 (2006)
  [hep-ph/0607170].

  
  

\bibitem{Buchmuller:2012ex} 
  W.~Buchm?ller, V.~Domcke and K.~Schmitz,
  JCAP {\bf 1304}, 019 (2013)
  [arXiv:1210.4105 [hep-ph]].

\bibitem{Buchmuller:2013zfa} 
  W.~Buchmuller, V.~Domcke and K.~Kamada,
  Phys.\ Lett.\ B {\bf 726}, 467 (2013)
  [arXiv:1306.3471 [hep-th]].

\bibitem{Buchmuller:2014rfa} 
  W.~Buchmuller, V.~Domcke and K.~Schmitz,
  JCAP11(2014)006
  [arXiv:1406.6300 [hep-ph]].
  

    

\bibitem{Domcke:2014zqa} 
  V.~Domcke, K.~Schmitz and T.~T.~Yanagida,
  Nucl.\ Phys.\ B {\bf 891}, 230 (2015)
  [arXiv:1410.4641 [hep-th]].



\bibitem{Battye:2010hg} 
  R.~Battye, B.~Garbrecht and A.~Moss,
  Phys.\ Rev.\ D {\bf 81}, 123512 (2010)
  [arXiv:1001.0769 [astro-ph.CO]].

  
\bibitem{Mukaida:2012qn} 
  K.~Mukaida and K.~Nakayama,
  JCAP {\bf 1301}, 017 (2013)
  [JCAP {\bf 1301}, 017 (2013)]
  [arXiv:1208.3399 [hep-ph]].

\bibitem{Harvey:1990qw} 
  J.~A.~Harvey and M.~S.~Turner,
  Phys.\ Rev.\ D {\bf 42}, 3344 (1990).
  
  
  
  
\bibitem{Coleman}
S.~Coleman,
Nucl.\ Phys.\ {\bf B262} (1985) 263.
  
  
  \bibitem{Qsusy}
A.~Kusenko,
Phys.\ Lett.\ {\bf B405} (1997) 108.

  
  
\bibitem{KuSh}
  A.~Kusenko and M.~E.~Shaposhnikov,
  Phys.\ Lett.\  B {\bf 418}, 46 (1998).
  
\bibitem{EnMc}
  K.~Enqvist and J.~McDonald,
  Phys.\ Lett.\  B {\bf 425}, 309 (1998); 
%
  Nucl.\ Phys.\  B {\bf 538}, 321 (1999).
  
\bibitem{KK1}
  S.~Kasuya and M.~Kawasaki,
  Phys.\ Rev.\  D {\bf 61}, 041301(R) (2000).

\bibitem{KK2}
  S.~Kasuya and M.~Kawasaki,
  Phys.\ Rev.\  D {\bf 62}, 023512 (2000).

\bibitem{KK3}
  S.~Kasuya and M.~Kawasaki,
  Phys.\ Rev.\  D {\bf 64}, 123515 (2001).
  
  
\bibitem{Higaki:2014eda} 
  T.~Higaki, K.~Nakayama, K.~Saikawa, T.~Takahashi and M.~Yamaguchi,
  Phys.\ Rev.\ D {\bf 90}, no. 4, 045001 (2014)
  [arXiv:1404.5796 [hep-ph]].
  
    
\bibitem{Fujii:2001xp} 
  M.~Fujii and K.~Hamaguchi,
  Phys.\ Lett.\ B {\bf 525}, 143 (2002)
  [hep-ph/0110072].
  
\bibitem{Fujii:2002kr} 
  M.~Fujii and K.~Hamaguchi,
  Phys.\ Rev.\ D {\bf 66}, 083501 (2002)
  [hep-ph/0205044].
  
\bibitem{Fujii:2002aj} 
  M.~Fujii and T.~Yanagida,
  Phys.\ Lett.\ B {\bf 542}, 80 (2002)
  [hep-ph/0206066].
  
  
\bibitem{Roszkowski:2006kw} 
  L.~Roszkowski and O.~Seto,
  Phys.\ Rev.\ Lett.\  {\bf 98}, 161304 (2007)
  [hep-ph/0608013].
  
  
\bibitem{Seto:2007ym} 
  O.~Seto and M.~Yamaguchi,
  Phys.\ Rev.\ D {\bf 75}, 123506 (2007)
  [arXiv:0704.0510 [hep-ph]].
  
  
  
\bibitem{ShKu}
  I.~M.~Shoemaker and A.~Kusenko,
  Phys.\ Rev.\  D {\bf 80}, 075021 (2009).
  
\bibitem{Doddato:2011fz} 
  F.~Doddato and J.~McDonald,
  JCAP {\bf 1106}, 008 (2011)
  [arXiv:1101.5328 [hep-ph]].
   
  
\bibitem{Kasuya:2011ix} 
  S.~Kasuya and M.~Kawasaki,
  Phys.\ Rev.\ D {\bf 84}, 123528 (2011)
  [arXiv:1107.0403 [hep-ph]].


\bibitem{Doddato:2012ja} 
  F.~Doddato and J.~McDonald,
  JCAP {\bf 1307}, 004 (2013)
  [arXiv:1211.1892 [hep-ph]].
  
  

\bibitem{Kasuya:2012mh} 
  S.~Kasuya, M.~Kawasaki and M.~Yamada,
  Phys.\ Lett.\ B {\bf 726}, 1 (2013)
  [arXiv:1211.4743 [hep-ph]].
  
  
\bibitem{KKY} 
  A.~Kamada, M.~Kawasaki and M.~Yamada,
  Phys.\ Lett.\ B {\bf 719}, 9 (2013)
  [arXiv:1211.6813 [hep-ph]].
  
\bibitem{Kamada:2014ada} 
  A.~Kamada, M.~Kawasaki and M.~Yamada,
  arXiv:1405.6577 [hep-ph].
  
      
    
    
\bibitem{Kasuya:2014ofa} 
  S.~Kasuya and M.~Kawasaki,
  Phys.\ Rev.\ D {\bf 89}, no. 10, 103534 (2014)
  [arXiv:1402.4546 [hep-ph]].
    
\bibitem{Kasuya:2014bxa} 
  S.~Kasuya and M.~Kawasaki,
  arXiv:1408.1176 [hep-ph].
    
\bibitem{Kasuya:2015uka} 
  S.~Kasuya, M.~Kawasaki and T.~T.~Yanagida,
  arXiv:1502.00715 [hep-ph].
  
    
\bibitem{Hiramatsu:2010dx} 
  T.~Hiramatsu, M.~Kawasaki and F.~Takahashi,
  JCAP {\bf 1006}, 008 (2010)
  [arXiv:1003.1779 [hep-ph]].
  
  
  
  
\bibitem{evap}
  A.~G.~Cohen, S.~R.~Coleman, H.~Georgi and A.~Manohar,
  Nucl.\ Phys.\  B {\bf 272}, 301 (1986).
  

  
\bibitem{KY} 
  M.~Kawasaki and M.~Yamada,
  Phys.\ Rev.\ D {\bf 87}, 023517 (2013)
  [arXiv:1209.5781 [hep-ph]].

    
  
\bibitem{Laine:1998rg} 
  M.~Laine and M.~E.~Shaposhnikov,
  Nucl.\ Phys.\ B {\bf 532}, 376 (1998)
  [hep-ph/9804237].
  
\bibitem{Banerjee:2000mb} 
  R.~Banerjee and K.~Jedamzik,
  Phys.\ Lett.\ B {\bf 484}, 278 (2000)
  [hep-ph/0005031].
  
    
  
  
\bibitem{ArkaniHamed:2004fb} 
  N.~Arkani-Hamed and S.~Dimopoulos,
  JHEP {\bf 0506}, 073 (2005)
  [hep-th/0405159].
  
\bibitem{Giudice:2004tc} 
  G.~F.~Giudice and A.~Romanino,
  Nucl.\ Phys.\ B {\bf 699}, 65 (2004)
  [Erratum-ibid.\ B {\bf 706}, 65 (2005)]
  [hep-ph/0406088].
  
  
\bibitem{ArkaniHamed:2004yi} 
  N.~Arkani-Hamed, S.~Dimopoulos, G.~F.~Giudice and A.~Romanino,
  Nucl.\ Phys.\ B {\bf 709}, 3 (2005)
  [hep-ph/0409232].
  
  
\bibitem{Wells:2004di} 
  J.~D.~Wells,
  Phys.\ Rev.\ D {\bf 71}, 015013 (2005)
  [hep-ph/0411041].
  
\bibitem{Hall:2011jd} 
  L.~J.~Hall and Y.~Nomura,
  JHEP {\bf 1201}, 082 (2012)
  [arXiv:1111.4519 [hep-ph]].
  
  
\bibitem{Ibe:2011aa} 
  M.~Ibe and T.~T.~Yanagida,
  Phys.\ Lett.\ B {\bf 709}, 374 (2012)
  [arXiv:1112.2462 [hep-ph]].
  
\bibitem{Ibe:2012hu} 
  M.~Ibe, S.~Matsumoto and T.~T.~Yanagida,
  Phys.\ Rev.\ D {\bf 85}, 095011 (2012)
  [arXiv:1202.2253 [hep-ph]].
  
\bibitem{Randall:1998uk} 
  L.~Randall and R.~Sundrum,
  Nucl.\ Phys.\ B {\bf 557}, 79 (1999)
  [hep-th/9810155].
  
\bibitem{Giudice:1998xp} 
  G.~F.~Giudice, M.~A.~Luty, H.~Murayama and R.~Rattazzi,
  JHEP {\bf 9812}, 027 (1998)
  [hep-ph/9810442].
  
\bibitem{Gherghetta:1999sw} 
  T.~Gherghetta, G.~F.~Giudice and J.~D.~Wells,
  Nucl.\ Phys.\ B {\bf 559}, 27 (1999)
  [hep-ph/9904378].
  
  
\bibitem{Kallosh:2006dv} 
  R.~Kallosh and A.~D.~Linde,
  JHEP {\bf 0702}, 002 (2007)
  [hep-th/0611183].
  
  
\bibitem{Izawa:1996pk} 
  K.~I.~Izawa and T.~Yanagida,
  Prog.\ Theor.\ Phys.\  {\bf 95}, 829 (1996)
  [hep-th/9602180].
  
\bibitem{Intriligator:1996pu} 
  K.~A.~Intriligator and S.~D.~Thomas,
  Nucl.\ Phys.\ B {\bf 473}, 121 (1996)
  [hep-th/9603158].
  
  

\bibitem{Buchmuller:2013uta} 
  W.~Buchm?ller, V.~Domcke and C.~Wieck,
  Phys.\ Lett.\ B {\bf 730}, 155 (2014)
  [arXiv:1309.3122 [hep-th]].




\bibitem{Nakayama:2012hy} 
  K.~Nakayama, F.~Takahashi and T.~T.~Yanagida,
  Phys.\ Lett.\ B {\bf 718}, 526 (2012)
  [arXiv:1209.2583 [hep-ph]].
  
  
\bibitem{Nakayama:2014xca} 
  K.~Nakayama, F.~Takahashi and T.~T.~Yanagida,
  Phys.\ Lett.\ B {\bf 734}, 358 (2014)
  [arXiv:1404.2472 [hep-ph]].

  
\bibitem{Bolz:2000fu} 
  M.~Bolz, A.~Brandenburg and W.~Buchmuller,
  Nucl.\ Phys.\ B {\bf 606}, 518 (2001)
  [Erratum-ibid.\ B {\bf 790}, 336 (2008)]
  [hep-ph/0012052].

\bibitem{Pradler:2006qh} 
  J.~Pradler and F.~D.~Steffen,
  Phys.\ Rev.\ D {\bf 75}, 023509 (2007)
  [hep-ph/0608344].
  
  
\bibitem{Kawasaki:1999na} 
  M.~Kawasaki, K.~Kohri and N.~Sugiyama,
  Phys.\ Rev.\ Lett.\  {\bf 82}, 4168 (1999)
  [astro-ph/9811437].
  
\bibitem{Kawasaki:2000en} 
  M.~Kawasaki, K.~Kohri and N.~Sugiyama,
  Phys.\ Rev.\ D {\bf 62}, 023506 (2000)
  [astro-ph/0002127].
  
\bibitem{Kawasaki:2004qu} 
  M.~Kawasaki, K.~Kohri and T.~Moroi,
  Phys.\ Rev.\ D {\bf 71}, 083502 (2005)
  [astro-ph/0408426].
  
  
\bibitem{Ichikawa:2005vw} 
  K.~Ichikawa, M.~Kawasaki and F.~Takahashi,
  Phys.\ Rev.\ D {\bf 72}, 043522 (2005)
  [astro-ph/0505395].
  
  
  
\bibitem{Kawasaki:2008qe} 
  M.~Kawasaki, K.~Kohri, T.~Moroi and A.~Yotsuyanagi,
  Phys.\ Rev.\ D {\bf 78}, 065011 (2008)
  [arXiv:0804.3745 [hep-ph]].

\bibitem{Hisano:2000dz} 
  J.~Hisano, K.~Kohri and M.~M.~Nojiri,
  Phys.\ Lett.\ B {\bf 505}, 169 (2001)
  [hep-ph/0011216].

\bibitem{Arcadi:2011ev} 
  G.~Arcadi and P.~Ullio,
  Phys.\ Rev.\ D {\bf 84}, 043520 (2011)
  [arXiv:1104.3591 [hep-ph]].

\bibitem{Ibe:2012hr} 
  M.~Ibe, A.~Kamada and S.~Matsumoto,
  Phys.\ Rev.\ D {\bf 87}, no. 6, 063511 (2013)
  [arXiv:1210.0191 [hep-ph]].
  
  


\bibitem{Bhattacherjee:2014dya} 
  B.~Bhattacherjee, M.~Ibe, K.~Ichikawa, S.~Matsumoto and K.~Nishiyama,
  JHEP {\bf 1407}, 080 (2014)
  [arXiv:1405.4914 [hep-ph]].
  
\bibitem{Fan:2013faa} 
  J.~Fan and M.~Reece,
  JHEP {\bf 1310}, 124 (2013)
  [arXiv:1307.4400 [hep-ph]].
  
  
\bibitem{Hisano:2006nn} 
  J.~Hisano, S.~Matsumoto, M.~Nagai, O.~Saito and M.~Senami,
  Phys.\ Lett.\ B {\bf 646}, 34 (2007)
  [hep-ph/0610249].

\bibitem{Cirelli:2007xd} 
  M.~Cirelli, A.~Strumia and M.~Tamburini,
  Nucl.\ Phys.\ B {\bf 787}, 152 (2007)
  [arXiv:0706.4071 [hep-ph]].
  
  
\end{thebibliography}
\end{document}